\documentclass[aps,prb,superscriptaddress,prb,floatfix,twocolumn,amsmath,amssymb,showpacs]{revtex4}
\usepackage{graphicx}
\usepackage{bm}

\begin{document}

\title{From the 2D graphene honeycomb lattice to 1D nanoribbons: dimensional crossover 
signals in the structural thermal fluctuations 
}

\author{S. Costamagna}
\affiliation{International School for Advanced Studies (SISSA),
Via Bonomea 265, I-34136 Trieste, Italy.}
\affiliation{
Facultad de Ciencias Exactas Ingenier{\'\i}a y Agrimensura, Universidad Nacional de Rosario and Instituto de
F\'{\i}sica Rosario, Bv. 27 de Febrero 210 bis, 2000 Rosario,
Argentina.}

\author{A. Dobry}
\affiliation{
Facultad de Ciencias Exactas Ingenier{\'\i}a y Agrimensura, Universidad Nacional de Rosario and Instituto de
F\'{\i}sica Rosario, Bv. 27 de Febrero 210 bis, 2000 Rosario,
Argentina.}

\date{\today}


\begin{abstract}
We study the dimensional crossover from 2D to 1D type behavior,
which takes place in the thermal excited  rippling of a
graphene honeycomb lattice, when one of the dimensions of the layer is reduced.
Through a joint study, by Monte Carlo (MC) atomistic simulations using
a quasi-harmonic potential and analytical calculations, 
we find that the normal-normal correlation function  
does not change its power law behavior in the long wavelength limit. 
However the system size dependency of the square of out of plane displacement 
$ \langle h^2  \rangle$ changes its scaling behavior when going from a 
layer to a nanoribbon. We show that a new scaling law
appears which corresponds to a truly 1D behavior and we estimate
the ratio of the sample dimensions where the crossover takes place as 
$R_{2D \leftrightarrow 1D}\approx 1.609$.
Having explored a wide number of realistic systems sizes,  
we conclude that narrow ribbons present stronger corrugations
than the square graphene sheets and we discuss the implications  
for the electronic properties of freestanding graphene systems.
\end{abstract}



\maketitle

{\textit{Introduction}}.
\label{intro}
The very existence of graphene\cite{Novoselov}, a first truly two dimensional crystal,  
is surprising because crystalline order is not expected to be possible in a two dimensional world.
This is a consequence of a Mermin-Wagner theorem \cite{MerminW} which forbids the ability of any continuous symmetry breaking, 
the translational one in this case, in two dimensions. However, graphene exists because the atoms can explore the other dimension
 by moving perpendicular to the sheet direction. In fact, the stability of a two-dimensional crystalline membrane embedded in a 
three dimensional world has been studied extensively in connection to biological membranes
\cite{NelsonPiranWeinberg}.
It has been concluded that the anharmonic coupling between the in plane and the out of plane phonons stabilizes an asymptotically flat phase but strongly corrugated by the presence of ripples, whose manifestation is the divergence of the mean square of the displacement perpendicular to the sheet in the thermodynamical limit.
The recently experimental demonstration of the existence of intrinsic ripples\cite{Meyer} in graphene has revived this subject of study. 
Moreover, numerical simulations using a realistic C-C interatomic potential supports the adequacy of the scaling theory of membranes 
in the continuum medium approach  for graphene\cite{fas-nat,fass2,membraba11} 
The connection of these ripples and their fluctuations with the electronic transport properties is a subject of great interest nowadays\cite{Castro}.

In addition to graphene sheets, ribbons where electrons are confined in the nano scale 
in the transversal direction, has been the subject of a lot of interest. 
The reason is that the lateral confinement may induce the presence of a gap in the electronic spectrum 
which is essential for many future nanoelectronics applications. 
However, the current success for the synthesis of the semiconducting graphene-based
nanoribbons is still far below the expectations. 
Lithographic patterning of graphene into nanodimensions has difficulties in 
controlling the nanopattern size and edge qualities. It is expected that these difficulties 
will be overcome in the near future. Therefore prediction of the structural 
stability as well its interplay with the electronic transport of graphene nanoribbons is strongly desirable.

\begin{figure}[ht]
\vspace{0.1cm}
\includegraphics[width=0.24\textwidth]{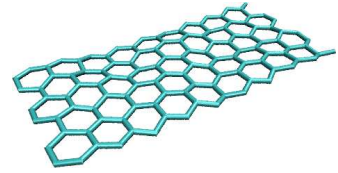}
\caption{
Esquematic view of the corrugated graphene honeycomb lattice. 
The dimensions of the flat sample are given by $L_x=3 l_x a_0$
and $L_y=\sqrt{3}l_y a_0$, where $a_0=1.42 \AA$ is the graphene lattice
constant and, $l_x$ ($l_y$) is the number of cells in the armchair(zigzag)-type edge. 
The total number of Carbon atoms is
$N=4 l_x l_y$. For this example $l_x=3$ and $l_y=10$ 
and hence $N=120$.}
\label{fig0}
\end{figure}

In this letter, we study the structural stability of graphene ribbons, 
extending the previous analysis conducted for the case of a square layer\cite{fass2}. 
From the elasticity theory and using the Self Consistent Screening Approximation (SCSA) 
to account for the coupling between in and out of plane phonons, 
we analyze the dimensional crossover which takes place when one of the dimensions of a layer is reduced. 
The normal-normal correlation function retains its power law behavior in the long wavelength limit. 
However, the mean square of the height fluctuation diverges 
with the length of the ribbons signaling an instability for the case of thin ribbons. 
We check this prediction by numerical simulations with the Montecarlo method using 
a phenomenological quasiharmonic potential. We conclude that ribbons should be strongly more rippled as compared with layer.   
 
\vspace{0.4cm}


{\textit{Stability condition from the elasticity theory}}.
\label{sec1}
Let $h(x,y)$ be the measure of the out-of-plane displacement of a particle sited
at the $(x,y)$ point, in a flat configuration, of a membrane. Then, 
the total bending energy is given by
\begin{eqnarray}
E_0=\int_{-\frac{L_x}{2}}^{\frac{L_x}{2}} dx \int_{\frac{L_y}{2}}^{\frac{L_y}{2}} dy \left[ \kappa \left(\nabla^2 h(x,y)\right)^2 \right]
\label{eq1}
\end{eqnarray}
\noindent where $\kappa$ is the bending rigidity. 
We are interested in comparing the case $L_x=L_y$ which 
represents a 2D membrane with the one $L_y>>L_x$ corresponding to a 
ribbon running in the y-direction.  
It is convenient to bring eq.(\ref{eq1}) to the momentum space formulation     
\begin{eqnarray}
E_0={\sum_{\textbf{k}=-(\Lambda_x,\Lambda_y)}^{(\Lambda_x,\Lambda_y)}}'(\left|\textbf{k}\right|^2)^2 \left|h(\textbf{k})\right|^2
\label{E0k}
\end{eqnarray}
where $\textbf{k}=(k_x,k_y)$ is the two dimensional momentum vector given as $k_i=\frac{2 \pi}{L_i}n_i$, 
$i=x,y$ in terms of the integer summation indexes $n_i$ 
. Note that this supposes periodic boundary condition (PBC) in both directions. We will also use PBC in the
 Montecarlo simulations we will show in the next Section. Therefore,  we take into account only the bulk excitations
 within the ribbon. 
The values of the width that make the ribbon unstable toward the thermally excitated rippling should be taken as an upper limit of stability.
Edge excitations, which are not taken into account in our approach, could destroy even thicker ribbons than which we consider here.  
In eq. (\ref{E0k}) $\Lambda_x$($\Lambda_y$) is the short distance ultraviolet cutoff of the order of the 
inverse of the lattice constants $a_x$($a_y$). 
The prime over the summation sign 
indicates that the $(0,0)$ point should be excluded because it corresponds to a rigid translation.   
The size dependency we are interested in will be related to
this low momenta limit that appears in all the k-summation.

The stability of this system can be analyzed according to the size dependency of $<h^2>$
the mean square of the height and $<\theta^2>\approx<|\nabla h|^2>$ the one of the bending angle
formed by the normal to the surface at a point with the global z-axis. 
For the isotropic membrane ($L=L_x=L_y$) 
we obtain for these quantities
\begin{equation}
<h^2>=\frac{1}{4\pi}\int_{\frac{2 \pi}{L}}^\Lambda dk k H(k) \sim L^2, 
\ \ \ 
<\theta^2>\sim \log L
\label{menhth}
\end{equation}   
where we have introduced the height-height correlation function given by
$H(k)=<h(\textbf{k})h(-\textbf{k})>=\frac{1}{2\beta\kappa k^4}$.
%
%

Note that this harmonic treatment predicts a logarithmic divergence with the 
system size which is interpreted as a crumpled instability. Moreover, 
as we discuss latter this instability is cured by anhamonic interactions with the in plane phonons.
 Before this, let us compare with the situation of a
 thin ribbon ($L_y>>L_x$). In this case the spectrum of the bending phonons 
$\omega_{ben}^2=\kappa \textbf{k}^4=\kappa \left[\left(\frac{2\pi}{L_x}n_x\right)^2+\left(\frac{2\pi}{L_y}n_y \right)^2\right]^2$
 is dominated by the $n_x=0$ term.  Equation (\ref{menhth}) now takes the form
\begin{equation}
<h^2>=\frac{1}{4 L_x \pi \beta\kappa}\int_{\frac{2 \pi}{L_y}}^{\Lambda_y} \frac{dk_y}{k_y^4}\sim \frac{L_y^3}{L_x}, 
\ \ \ 
<\theta^2>\sim \frac{L_y}{L_x}
\label{h2t2freerib}
\end{equation}  
The instability here is of course more severe that in the 2D case. 
This is the first indication that thin ribbons are much more sensitive to thermal out plane fluctuation than a square membrane. 

However, the bending modes are coupled to the in-plane ones. A simple way to include this coupling is by taking into account that 
in the elastic two dimensional theory given by the energy 
\begin{eqnarray}
E_{pl}=\int_{-\frac{L_x}{2}}^{\frac{L_x}{2}} dx \int_{\frac{L_y}{2}}^{\frac{L_y}{2}} dy \left[ 2 \mu u^2_{i j}+\lambda  u^2_{i i} \right]
\end{eqnarray}
the strain tensor $u_{ij}$ should include a quadratic term in $h$
($\lambda$ is the lame constant and $\mu$ the the shear modulus). 
Their lowest order is therefore given by
\begin{eqnarray}
u_{ij}=\frac12\left[\left(\partial_ju_i+\partial_i u_j \right)+\partial_i h \partial_j h\right] 
\end{eqnarray}\
$u_i$ are the $x$ and $y$ Cartesian components of the displacements from the equilibrium situation. 

The simplest but otherwise precise treatment as compared with the Montecarlo 
numerical simulation\cite{fass2} is the Self Consistent Screening Approximation 
(SCSA) of LeDoussal and Radzihovsky \cite{LR}. 
The in-plane phonons are integrated out in the partition function,  generating 
an effective quartic interaction between the out of plane modes. The correlation function $H(\textbf{k})$ 
acquires a self-energy correction $H^{-1}(\textbf{k})=\kappa  k^4+\Sigma(\textbf{k})$ which is self-consistently determined. 
The resulting coupled  integral equations could be analytically solved in the long wavelength limit under the assumption $H^{-1}\approx\Sigma(\textbf{k})=Z k^{4-\eta}$ $Z$ is a non universal
 amplitude and $\eta\approx0.821$ is obtained by solving analytically the integrals in terms of the $\Gamma$ function.
 A crucial point for the present analysis is that $\eta$ cannot not depend on the system size (at least for big enough systems),
 because the integral in $k$ involved in its calculation is convergent with the low momenta infrared limit. 
Therefore, we expect that $\eta$ will not change when we go from a square membrane to a ribbon. 
This will be checked by Monte Carlo(MC) simulations.

Coming back to the membrane and including the renormalization of $H^{-1}(\textbf{k})$ given  by the SCHA, Eq. (\ref{menhth}) becomes
\begin{equation}
<h^2>=\frac{1}{4\pi \beta\kappa}\int_{\frac{2 \pi}{L}}^\Lambda \frac{dk}{k^{3-\eta}}\sim L^{2-\eta},
\ \ \ 
<\theta^2>\sim  L^{-\eta}
\label{t2renmem2}
\end{equation}
The bending angle converges for $L\rightarrow\infty$ implying a long range order on the normals. However  $<h^2>$ diverges
which is interpreted as the origin of a strong rippling of the membrane which in 
fact observed in suspended graphene samples.

\begin{figure}[hbt]
\includegraphics[width=0.46\textwidth]{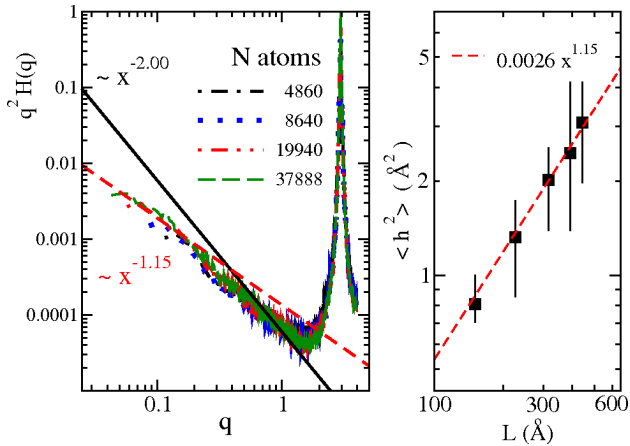}
\caption{(a) (Color online)
$q^2 H(q)/N$ against $q$ at T = 300 K for different 2D system sizes 
as indicated on the plot. 
The dashed lines show the asymptotic behaviors. 
(b) $ \langle h^2 \rangle $ as a function of the average
linear system size $\tilde{L}=\sqrt{L_xL_y}$ compared to the scaling 
law $ \langle h^2 \rangle=C_{2D} L^{2-\eta}$ with $C_{2D}=0.0026$
and $\eta=0.85$ (dashed line). Both axes are in logarithmic scale.
This Figure has to be compared with the figs. 2 and 3 of Ref. \onlinecite{fass2}.
}
\label{fig11}
\end{figure}
 
{\textit{Monte Carlo simulations for the honeycomb lattice.}}
The scaling laws described above have been obtained by analyzing the
asymptotic behavior due to thermal fluctuations of continuous membranes.
Recently, its validity was proven on a discrete 2D
lattice, the graphene layer\cite{fass2}.
This was done with MC simulations using a potential
suited for the Carbon compounds by adopting PBC on square samples.
In fig. \ref{fig11} we reproduce these results as a previous step       
before the study of the ribbons. Figure \ref{fig11} (a) shows $q^2 H(q)/N$
, which is the normal-normal correlation function because the unit vector normal to the surface $\textbf{n}$
is connected with $h$ by 
$\textbf{n}\approx\textbf{e}_z-\nabla h$. It is 
calculated over 2D layers of several sizes as indicated on the plot.
Besides this, fig. \ref{fig11}(b) shows $ \langle h^2\rangle $   
as a function of the average linear system size $\tilde{L}=\sqrt{L_xL_y}$
compared to the eq. (\ref{t2renmem2}).
These results were obtained through  ordinary MC simulations
including the wave-moves and using the quasi-harmonic (QH) potential 
introduced in Ref. \onlinecite{fass2}.
The details of the calculation, including the implementation of the wave-moves,
follow the lines described in this reference.   
We have found the right acceptation range in the simulations
by using amplitudes for wave-moves of the type $A=C_{wm}/\tilde{L}^{R_{scal}}$, 
with $C_{wm}=0.01$ and $R_{scal}=0.32$. The observed agreement 
should be taken as a check of the reliability of our sampling procedure.

\begin{figure}[hbt]
\includegraphics[width=0.475\textwidth]{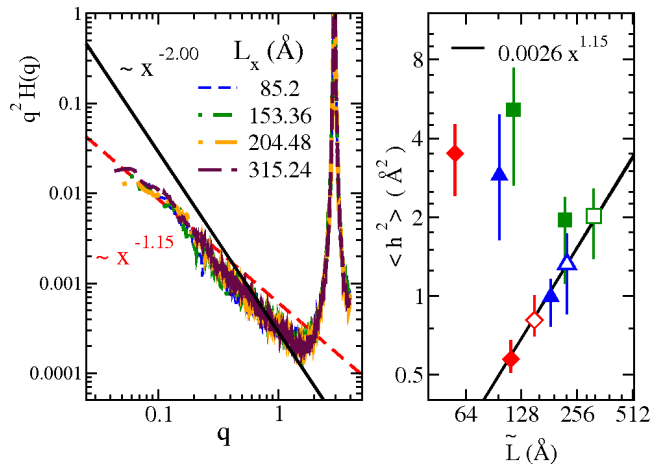}
\caption{(a) (Color online)
$q^2 H(q)/N$ against $q$ at T = 300 K 
calculated for different ribbon widths $L_x$, starting from the
2D $N=37888$ ($L_x=315.24 \AA$)layer, as indicated on the plot.
The asymptotic behavior remains valid
independently of the ribbon width as was discussed previously in the text .
(b) $ \langle h^2 \rangle$ as a function of the average
linear system size $\tilde{L}=\sqrt{L_xL_y}$ compared to the scaling 
law $\langle h^2 \rangle=C_{2D} L^{2-\eta}$ used in fig.\ref{fig11} (b).
Open symbols correspond to square lattices, being square for $N=37888$, triangle for $N=19440 $ and diamond for $N=8640$. 
Filled symbols represent ribbon type lattices obtained from the square ones by reducing the $L_x$ dimension.
Starting from each 2D layer, in all cases the scaling law is broken at a certain value 
that we analyze further below. Both axes are in logarithmic scale.}
\label{fig22}
\end{figure}

From the previous section it is expected that when one of the dimensions of the 
square sample is reduced a distinct behavior should appear  
in the scaling of $ \langle h^2 \rangle $, otherwise, 
$q^2 H(q)/N$ should show the same scaling. 
Figure \ref{fig22} (a) shows that the asymptotic behavior of $q^2 H(q)/N$ remains valid
starting from the square sample with N=37888 by reducing $L_x$ as indicated
(the same behavior is obtained by decreasing $L_y$).
In fig. \ref{fig22} (b) it can be observed that the scaling behavior
of $ \langle h^2 \rangle $ against $\tilde{L}$, corresponding to three different samples, 
is followed only for certain widths but it is not fulfilled when the sample 
becomes narrow enough. In all the cases we have verified that the same 
behavior is obtained by decreasing $L_y$.

As previously discussed we have adopted PBC in all the simulations. 
In this form we avoid edge reconstruction effects
on the samples which could lead to unwanted effects for our study.
We have limited the width of the 
studied ribbons to be large enough values in a way that bulk effects accounted by PBC are an  important 
part of the physics of this problem.

\vspace{0.4cm}

{\textit{The one-dimensional behavior}}.
\label{oneD}
From fig. \ref{fig22} (a) it can be established 
that the value of $\eta$ does not change when going from the square membrane to a ribbon. 
Now, a similar analysis leading to Eq. \ref{t2renmem2} 
 can be undertaken for the thin ribbons. 
In this case, the mean square fluctuations acquire
a dependence toward the long direction $L_y$ and the scaling laws become 
\begin{equation}
<h^2>\sim \frac{L_y^{3-\eta}}{L_x},
\ \ \ \ 
<\theta^2>\sim  \frac{L_y^{1-\eta}}{L_x}
\label{t2renrib2}
\end{equation} 

\begin{figure}[ht]
\vspace{0.2cm}
\includegraphics[width=0.4\textwidth]{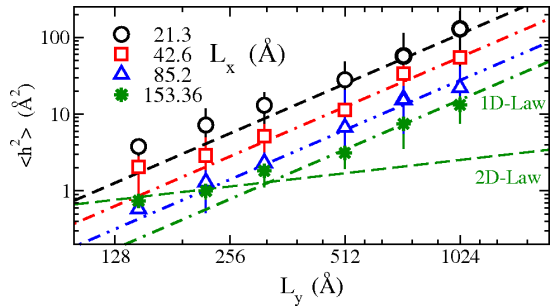}
\caption{(Color online)
$\langle h^2 \rangle$ versus the longitude of the ribbon $L_y$
for several ribbons width $L_x$ as indicated.
The dotted line serves to compared to the 
scaling law $\langle \langle h^2 \rangle \rangle=(C_{1D}/L_x) L_y^{3-\eta}$
with $C_{1D}=0.0055$. The dashed line represent the 2D scaling for the 
widest ribbon (see the text). The crossover 2D to 1D can be seen clearly in this case. 
Both axis are in logarithmic scale. 
}
\label{fig33}
\end{figure}

The main result of this work is present in fig. \ref{fig33}
where it can be observed the scaling of $\langle h^2 \rangle$ against $L_y$
measured with the MC simulations for different ribbon widths as indicated.
The comparison with the eq. (\ref{t2renrib2}) using the law $(C_{1D}/L_x) L_y^{3-\eta}$
with $C_{1D} = 0.0055$ shows an excellent agreement in all the cases 
for the larger values of $L_y$ where the 1D aspect is stronger. 
The lowest $L_y$ point of the $L_x=153.36\ \AA$ sample (star symbol)
corresponds to the 2D lattice N=37888 where, as we already discussed, 
the expectation value of fluctuations is described by Eq. (\ref{t2renmem2}).
For this reason we also include on the plot this law using  
the fitted value $C_{2D}=0.0026$, and $L_x=153.36 \AA$. This point 
is the only one included in this figure which corresponds to a 2D layer. 
The change of the scaling, thus, emerges clearly and the fittings allow us,
by equaling both laws, Eq. (\ref{t2renrib2}) (1D) and Eq. (\ref{t2renmem2}) (2D), 
to estimate the ratio $R_{2D \leftrightarrow 1D}$ in which the crossover 
should take place as 
\begin{equation}
R_{2D \leftrightarrow 1D}=\frac{L_y}{L_x}=
\frac{C_{1D}}{C_{2D}}^{2/(\eta-4)}
\approx 1.609
\end{equation}

{\textit{Conclusions}}.
\label{conclusions} 
The long range order on the ribbons is not restored by the inclusion of $\eta$ since eq. (\ref{t2renrib2})
implies that $<\theta^2>$ remains unbounded in the thermodynamical limit and
therefore the free thin ribbons are not stable against a crumpled transition.
For the finite samples studied here this transition has not appeared.
In the experimental situation, the free standing samples are always
supported at the edges, hence the induced tension can provide room
for a higher stabilization of the ribbons\cite{belgica3}.

As the QH potential employed here is softer than those ones best 
suited for graphene\cite{rebo,L2}, to extend our conclusions to the graphene 
nanoribbons we emphasize that,  $R_{2D \leftrightarrow 1D}$ could be greater
than the estimated here. However we do not expect a qualitative change in the overall behavior which is related
to general scaling arguments. Even more, the use of more realistic Carbon-Carbon potentials would be desirable 
in the studies where the edge reconstruction is taken into account in addition to thermal excited ripples
or the more stable configurations of low number of Carbon atoms\cite{belgica2}. 
Finally, we remark that $R_{2D \leftrightarrow 1D}$ do not represent the limit 
in which the crumpled instability would disassemble the lattice, 
but instead, where the crossover  
takes place followed by the change on the scaling laws described here. 
A strong corrugation could imply big changes on the electronic spectra 
of the nanorribons and may even result in gap openings on the electronic spectra\cite{seba}.
\vspace{0.2cm}

{\textit{Acknowledgments}}.
\label{agradecimientos}
We thank D. Mastrogiusseppe for carefully reading our manuscript  
and A. Fasolino for helpful comments.
This work was partially
supported by PIP  11220090100392 of CONICET, and
PICT 1647  and PICT R 1776 of the ANPCyT.



\end{document}